
\input harvmac

\def\Title#1#2{\rightline{#1}\ifx\answ\bigans\nopagenumbers\pageno0
\vskip0.5in
\else\pageno1\vskip.5in\fi \centerline{\titlefont #2}\vskip .3in}

\font\caps=cmcsc10

\noblackbox
\parskip=1.5mm


\def\npb#1#2#3{{\it Nucl. Phys.} {\bf B#1} (#2) #3 }
\def\plb#1#2#3{{\it Phys. Lett.} {\bf B#1} (#2) #3 }
\def\prd#1#2#3{{\it Phys. Rev. } {\bf D#1} (#2) #3 }

\def\bb#1{{\tt hep-th/#1}}




\lref\rwitt{E. Witten, {\it ``On S-duality in Abelian Gauge Theory"},
IAS preprint IASSNS-HEP-95-36, \bb{9505186.}}
\lref\rver{E. Verlinde, {\it ``Global Aspects of Electric-Magnetic
Duality"}, CERN preprint CERN-TH/95-146, \bb{9506011}}
\lref\rspwil{J. Cardy and E. Rabinovici, \npb {205}{1982}{1\semi}
J. Cardy, \npb {205}{1982}{17\semi}
A. Shapere and F. Wilczek, \npb {320}{1989}{669.}}

\lref\rrocver{M. Ro{\v c}ek and E. Verlinde, \npb {373}{1992}{630.} }
\lref\rbuscher{T.H. Buscher, \plb {201}{1988}{466 \semi}
A. Giveon, M. Porrati and E. Ravinovici, Phys. Rept. {\bf 244} (1994)
77.}
\lref\rsenschw{J. Schwarz and A. Sen, \npb {411}{1994}{35.}}
\lref\rgrosskleb{D.J. Gross and I. Klebanov, \npb {344}{1990}{475.}}
\lref\rwit{E. Witten, \plb {86}{1979}{283.}}
\lref\rib{A. Font, L. Iba{\~n}ez, D. L{\"u}st and F. Quevedo, \plb
{249}{1990}{35. \semi}
S. J. Rey, \prd {43}{1991}{526.\semi}
A. Sen, Int. J. Mod. Phys. {\bf A9} (1994) 3707.}
\lref\rquevedo{X.C. De la Ossa and F. Quevedo, \npb {403}{1993}{337.}}
\lref\rwitvafa{C. Vafa and E. Witten, \npb {432}{1994}{3.}}
\lref\rseiwitint{N. Seiberg and E. Witten, \npb {426}{1994}{19. \semi}
K. Intriligator and N. Seiberg, \npb {431}{1994}{551.}}
\lref\rseidos{N. Seiberg, \npb {435}{1995}{129.}}
\lref\rmonol{C. Montonen and D. Olive, \plb {72}{1977}{117.}  }
\lref\rdimred{J. Harvey, G. Moore and A. Strominger, {\it ``Reducing
S-duality to T-duality"} , EFI-95-01, YCTP-P2-95, \bb{9501022. \semi}
M. Bershadsky, A. Johansen, V. Sadov and C. Vafa {\it ``Topological
reduction of 4-d SYM to 2-d sigma models"}, Harvard preprint,
HUTP-95-A004, \bb{9501096.}}
\lref\rporrati{L. Girardello, A. Giveon, M. Porrati and A. Zaffaroni,
\plb {334}{1994}{331.}}



\line{\hfill PUPT-95-1543}
\line{\hfill {\tt hep-th/9506137}}
\vskip 1cm

\Title{\vbox{\baselineskip 12pt\hbox{}
 }}
{\vbox {\centerline{Generalized Abelian S-duality}
\medskip
\centerline{and coset constructions }  }}

\centerline{$\quad$ {\caps J. L. F. Barb\'on}}
\smallskip
\centerline{{\sl Joseph Henry Laboratories}}
\centerline{{\sl Princeton University}}
\centerline{{\sl Princeton, NJ 08544, U.S.A.}}
\centerline{{\tt barbon@puhep1.princeton.edu}}
\vskip 0.4in

 Electric-magnetic duality and higher dimensional analogues are obtained
as symmetries in generalized coset constructions, similar to the
axial-vector duality of two-dimensional coset models described by
Ro{\v c}ek and Verlinde. We also study global aspects of duality between
$p$-forms and $(d-p-2)$-forms in $d$-manifolds. In particular, the
modular duality anomaly is governed by the Euler character as in four
 and two dimensions. Duality transformations of Wilson line operator
insertions are also considered.


\Date{6/95}

\newsec{Introduction}
Recent exact results in supersymmetric gauge theories have prompted a
renewed interest in S-duality, understood as a generalization of
electric-magnetic weak-strong coupling duality. In addition to the
classic selfduality conjecture of Montonen and Olive in $N=4$
supersymmetric Yang-Mills
 \refs\rmonol , and its stringy generalization \refs\rib ,
many new examples and conjectures have been proposed recently, both in
quantum field theory and string theory. In some cases
\refs\rseiwitint , S-duality of a low energy abelian gauge theory
plays an important role in the solution of the infrared physics of
certain $N=1$ and $N=2$ gauge theories. A generalization of the
Montonen-Olive duality to $N=1$ Super-QCD in a non abelian Coulomb
phase was also proposed in ref. \refs\rseidos , with the striking
property that the number of gauge degrees of freedom is totally
different in the two dual descriptions.

 In this last example, and also in the Montonen-Olive case, the
duality is supposed to operate in the full non abelian theory, and not
only in a spontaneously broken phase with an abelian low energy
theory. Evidence for this fact in the $N=4$ theory has been presented
in \refs\rwitvafa\ and \refs\rporrati . On the other hand, the dual
gauge group in the Montonen-Olive sense si found by just dualizing the
Cartan subalgebra. Also, dimensional reduction to two dimensions
\refs\rdimred\ apparently projects non abelian S-duality onto standard
abelian T-duality.

Unlike the non abelian case, S-duality in abelian gauge theories can
be described quite explicitly by different methods involving just
gaussian path integrals. In view of the previous remarks, it is
interesting to understand abelian duality in its full generality, and
in particular the relations and analogies to the two dimensional
case. Recently, discussions of global aspects in electric-magnetic
duality have appeared in refs.\refs\rwitt\ and \refs\rver . In this
paper, we present a generalization of some of the results in these
papers to arbitrary dimensions. In particular, we determine the
modular duality anomaly for the general duality between $p$-forms and
$(d-p-2)$-forms and find the same result as in four dimensions, up to
sign factors.

In order to pursue further the analogies between higher dimensional
S-duality and two dimensional duality of sigma-models we generalize
the Ro{\v c}ek-Verlinde coset construction to the duality between
$p$-form gauge theories in even dimensions. This might be interesting
to address more complicated cases because coset constructions are
examples of redundant gauge symmetries (the gauge fields are infinitely
strongly coupled and do not propagate). The coset construction we
describe is similar to the work of \refs\rsenschw\ on duality
symmetric actions, although important differences are pointed out.

\newsec{Coset constructions}
\subsec{The two-dimensional case}

Abelian T-duality is a well known symmetry of string perturbation
theory. At the world-sheet level it is a non-perturbative
Kramers-Wannier transformation or, in continuum language, a Hodge
duality transformation: $\partial_{\alpha} \theta \rightarrow
\epsilon_{\alpha\beta} \partial^{\beta} \theta$. This mapping is
non-local in terms of the field $\theta$ (it introduces winding modes
and vortices) and exchanges equations of motion
$\partial_{\alpha}\partial^{\alpha} \theta =0$ with the Bianchi identity
$\epsilon^{\alpha\beta}\partial_{\alpha}\partial_{\beta} \theta=0$.
Presented in this form,  it
is  an analogue of four dimensional electric-magnetic duality which
does the same in terms of the vector potential $A_\mu$ :$ dA\rightarrow
*dA$. Indeed, the naive dimensional reduction of the Maxwell theory to
two dimensions yields two scalar fields as the internal components of
the photon, and the four dimensional S-duality is mapped into these
scalars as two dimensional T-duality, global topological
electric-magnetic fluxes are mapped into winding modes, and monopoles
yield two dimensional vortex configurations (winding modes around a
singular point).

The standard manipulation to exhibit two-dimensional Hodge duality
consists in writing the path integral in first order form, changing
variables from $\theta$ to $d\theta =A$. In doing so, we assume that, as a
sigma-model, there is a target space isometry under constant shifts of
the $\theta$ field: $\theta\rightarrow \theta + \xi$, so that the action is
of the form
\eqn\ssigma {
S(\theta) = {1\over 2\pi} \int d^2 z \, g_{\theta\theta} \partial
\theta
{\overline \partial}\theta +\cdots
 = {1\over 8\pi} \int g_{\theta\theta}
 (d\theta)^2
 + \cdots}
with $g_{\theta\theta}$ independent of $\theta$. The
 change of variables can be readily implemented in a lattice
regularization (see the nice discussion in \refs\rgrosskleb).
There is a local constraint $*dA=0$ which can be written as a
functional integral over a Lagrange multiplier field $\tilde\theta$ and
integrating out the one-form $A$ completes the proof.

A simple algorithm to keep track of the functional measures in the
continuum language uses a variant of the first order formalism (see
\refs\rbuscher\ , \refs\rrocver\ ), in which one gauges the isometry and
cancels the non-propagating gauge field by means of the same local
constraint $*dA=0$. In formulas
\eqn\truco
{
\eqalign{Z&= \int  D\theta\,\, {\rm e}^{-S(d\theta)}
 =
\int { D\theta  D A \over {\rm Vol}(G)}\, \delta(*dA)\,
{\rm e}^{ -S(d\theta +A)}
\cr
&= \int{ D\theta  DA  D {\tilde \theta} \over {\rm
Vol} (G)}\, {\rm e}^{-{i\over 2\pi} \int A\wedge d{\tilde \theta}
-S(d\theta +A)^2 }
}}
Since there was an isometry the action is quadratic in $A$
($g_{\theta\theta}$ still depends on other fields in general). Then we
may gauge fix $\theta =0$ and integrate out $A$ to get the dual
version of the model
\eqn\sigmadual {
Z= \int  D{\tilde \theta}\, {\rm exp}\left(- {1\over 8\pi} \int {1\over
g_{\theta\theta}} (d{\tilde \theta})^2\right)
}
with the characteristic $g_{\theta\theta} \rightarrow
1/  g_{\theta\theta}$ form.
A careful treatment of the ultralocal jacobians in the integration
measure yields in addition  a shift in the dilaton background field
$\Phi \rightarrow \Phi + {\rm log}( g_{\theta\theta})$. Formally, the
local measure is regularized preserving the following structure:
\eqn\measure
{ D\theta \sim \prod_z \left( {d\theta_z \over \sqrt{2\pi}  }
\sqrt{g_{\theta\theta} (z)}\right) }

The coupling between the gauge field and the Lagrange multiplier is
designed such that both the original field $\theta$ and its dual have
the same periods. If the original current has non-trivial holonomies
around homology 1-cycles
\eqn\holon
{\oint_{\gamma} d\theta \in 2\pi {\bf Z} }
then the gauged model must be invariant under large gauge
transformations $A\rightarrow A-d\epsilon$ where $d\epsilon$ has
$2\pi\times ({\rm integer})$ periods. This condition is met if and only
if $d{\tilde \theta}$ has the same periodicity, so that
$$
{\rm exp}\left({i\over 2\pi}
 \int d\epsilon \wedge d{\tilde\theta}\right) = {\rm e}^{2\pi i
n\cdot m} = 1
$$
In fact, formula \holon\ is an abuse of notation. One should think of
$d\theta$ as an exact 1-form plus a harmonic piece which is
responsible for the periods.

 An analogous procedure for four-dimensional electric-magnetic duality
was recently used by Witten in \refs\rwitt . There the ``isometry" to
be gauged is the shift of the vector potential
$$
A\rightarrow A+B
$$
by an arbitrary 1-form B. This is achieved introducing a 2-form
gauge field $G$, with a 3-form field strength $dG$, which is required
to vanish as a constraint. A convenient representation uses a one-form
Lagrange multiplier
$$
\delta[dG] \sim \int {D{\tilde A} \over {\rm Vol}({\tilde G})}
\,\,{\rm e}^{{i\over 2\pi} \int dG\wedge {\tilde A}}
$$
where the dual gauge symmetry ${\tilde G}$ takes care of the ambiguity
${\tilde A}\rightarrow {\tilde A} +d\phi$ in the exponentiation of the
delta functional. The occurrence of global electric-magnetic fluxes
wrapped around homologically non-trivial two-submanifolds exactly
parallels the previous two-dimensional case. In section 4 we will
exploit this technique to investigate the general duality relation
between $p$-forms and $(d-p-2)$-forms in an arbitrary $d$-manifold, as well as
to compute the non-local ``disorder operators" dual to generalized
Wilson $p$-lines.

The previous formalism treats the original and dual variables (the
Lagrange multipliers) in a rather asymmetric fashion. For example, the
dual photon $\tilde A$ only acquires a kinetic energy term after the
fake gauge field has been eliminated. A more symmetric procedure
exists for two dimensional T-duality, as explained by Ro{\v c}ek and
Verlinde in \refs\rrocver.
    Roughly speaking, the method constructs both
versions of the model as equivalent cosets of a single sigma model
with doubled degrees of freedom and carefully chosen couplings. The
duality symmetry corresponds to a discrete field redefinition in the
doubled theory (in the context of WZW conformal field theories it is a
Weyl transformation). To illustrate the duality between \ssigma\
and \sigmadual\ consider the following sigma model with two independent
fields $\theta_{LR}$
\eqn\LR
{S_{LR} = {1\over 2\pi} \int d^2 z (\partial \theta_L {\bar
\partial}\theta_L + \partial \theta_R {\bar \partial}\theta_R
+ 2B \partial \theta_R {\bar\partial}\theta_L + ...)}
where $B$ is independet of $\theta_L, \theta_R$ but may depend on
other fields of the sigma model. This action has a $U(1)_L \times
U(1)_R$ affine symmetry generated by the chiral currents
$$
J^L = \partial \theta_L + B \partial \theta_R  \,\,\,,\,\,\,\,
{\bar J}^R = {\bar \partial}\theta_R + B
{\bar\partial}\theta_L
$$
The axial-vector cosets are constructed by gauging with the minimal
coupling prescriptions
\eqn\mincop
{d\theta_R \rightarrow d\theta_R + {A\over 2} \,\,\,,\,\,\,\, d\theta_L
\rightarrow d\theta_L \pm {A\over 2}}
and the adition of a gauge invariant term
\eqn\extra
{S' = {1\over 4\pi} \int d^2 z (\theta_R \mp \theta_L) (\partial
{\bar A} - {\bar \partial}A)}

The gauge fields are kept non-propagating so that they simply project
out the physical Hilbert space. Introducing the axial-vector
combinations $\theta = \theta_R +\theta_L$, ${\tilde \theta} =
\theta_R - \theta_L$ and integrating out the gauge fields we find the
following semiclassical sigma models
$$
S_{\rm Vector} = {1\over 2\pi} \int d^2 z \left({1+B \over 1-B}\right)
\partial \theta {\bar\partial}\theta +...
$$
$$
S_{\rm Axial} = {1\over 2\pi} \int d^2 z \left({1-B\over 1+B}\right)
\partial{\tilde \theta} {\bar\partial}{\tilde \theta}
+...
$$
defining $g_{\theta\theta} = {1+B\over 1-B}$ we obtain both dual sigma
models by switching from vector to axial gauging, that is
  $B\rightarrow -B$.
This transformation can be undone by a field reparametrization
$\theta_L \rightarrow -\theta_L$ in the doubled action. An important
point  is that this transformation is not a classical symmetry
of the action \LR\ as it stands, but it is always a symmetry of the
path integral.

\subsec{Generalization to even p forms in 2p+2 dimensions}
The previous construction readily generalizes to $\it{even}$ $p$ forms
in $d= 2p+2$ dimensions.
Let us define a pair of $p$-form potentials $A_L, A_R$ with field
strengths $F_{LR} = d A_{LR}$, and the axial-vector combinations $F=
F_R + F_L$, ${\tilde F} = F_R - F_L$. Consider the doubled action
\foot{Our conventions for the product of forms in this section and the
 rest of
the paper are $\alpha_n \beta_n \equiv \alpha_n \wedge * \beta_n  =
{1\over n!} \alpha_{i_1 ...i_n}\beta^{i_1 ... i_n} {\rm d(Vol)}$}

\eqn\adre {
4\pi \,{\cal L}_{\rm doubled} = F_L^2 + F_R^2 + 2\mu\, F_R F_L +
 2i\mu\, F_R
*F_L
}

The coset construction is obtained by a direct extension of the formulas
\mincop\ with a $(p+1)$-form gauge field $G$
\eqn\mincopp
{F_R \rightarrow F_R + {G\over 2} \,\,\,,\,\,\, F_L \rightarrow F_L
\pm {G\over 2}}
and adding a gauge invariant coupling
\eqn\otroextra
{ \int {\cal L}' = {i\over 4\pi} \int dG *(A_R \mp A_L) = {i\over
4\pi} \int G*d(A_R \mp A_L) }
The vector gauging leads to the model
$$
4\pi\, {\cal L}_{\rm gauged} = {1+\mu \over 2} F^2 + {1-\mu \over 2}
({\tilde F} +G)^2 +i\mu\, ({\tilde F} +G)*F + iG*F
$$
The gauge field $G$ is kept non propagating (ie. extreme strong
coupling), so that
 we can integrate it out and gauge fix ${\tilde A} =0$ with the
result
\eqn\a
{4\pi\, {\cal L} = {1+\mu \over 1-\mu} F^2 \equiv {4\pi \over g^2} F^2 }

With this definition of the gauge coupling, duality amounts to $\mu
\rightarrow -\mu$ just as in the two-dimensional case. This is
precisely  provided by de axial gauging
$$
4\pi\, {\widetilde{\cal L}}_{\rm gauged}
 = {1+\mu \over 2} (F+G)^2 + {1-\mu \over 2}
{\tilde F}^2 +i\mu\, {\tilde F}*(F+G) + iG*{\tilde F}
$$
Proceeding as before, upon $G$ integration we arrive at the dual theory
\eqn\b
{4\pi\, {\widetilde{\cal L}} = {1-\mu \over 1+\mu} {\tilde F}^2 \equiv {4\pi
\over {\tilde g}^2} {\tilde F}^2 }
and ${\tilde g} = 4\pi /g$ as desired (we have chosen the
normalization of the coupling such that the duality is $g\rightarrow
4\pi /g$ in any dimension). In arriving at this result it is very
important that the field redefinition interchanging both gaugings: $F_L
\rightarrow -F_L $ or $F\leftrightarrow {\tilde F}$ is equivalent to
$\mu \rightarrow -\mu $ in the doubled theory. This is the case for
$p+1 = {\rm odd}$ thanks to the identity
$$
F_{p+1} * {\tilde F}_{p+1} = -{\tilde F}_{p+1} * F_{p+1}
$$
In dimension $d=2p+2$ with odd $p$ the wedge product of $(p+1)$ forms is
symmetric (that is the reason why we can define a $\theta$ angle). So
in particular the analogue of axial-vector duality does not work in
dimension four.

\subsec{Odd p forms in 2d+2 dimensions}

In order to further discuss the odd $p$ case it is useful to define
the selfdual and anti-selfdual projections $F^{\pm} = (F\pm *F)/2$.
Let us define the following doubled action:
\eqn\otradoble
{\eqalign{4\pi\,{\cal L}_{\rm doubled}& = {1+\mu\over 2}(F^+)^2 + {1+{\overline
\mu}\over 2} (F^-)^2 + {1-\mu\over 2}({\tilde F}^+)^2 +
{1-{\overline\mu} \over 2} ({\tilde F}^-)^2\cr &+i\mu\, F^+ {\tilde F}^+ -
i{\overline \mu}\,F^- {\tilde F}^- \cr} }
where now $\mu$ is a complex number, to allow for a non zero $\theta$
angle. If we gauge the $\tilde A$ field by the minimal prescription
${\tilde F}^{\pm} \rightarrow {\tilde F}^{\pm} + G^{\pm}$,  plus the
gauge invariant term
$$
\int {\cal L}' = {i\over 4\pi} \int dG * A
$$
we obtain, after gauge fixing and $G^{\pm}$ integration
\eqn\directa
{4\pi\, {\cal L} = {1+\mu \over 1-\mu}(F^+)^2 + {1+{\overline\mu}\over
1-{\overline\mu}} (F^-)^2 }
so that, defining $\tau = i {1+{\overline \mu} \over 1-{\overline \mu}}$
 we get the standard
lagrangian\foot{This definition is consistent with the usual
normalization $\tau = {4\pi i \over g^2 } + {\theta \over 2\pi}$.}
$$
{\cal L} = {i\over 4\pi} \left( {\overline\tau} (F^+)^2 - \tau (F^-)^2
\right)
$$
{}From this manipulation it is clear that duality $\tau \rightarrow
{\tilde \tau} = -1/\tau$ would again correspond to $\mu \rightarrow
-\mu$ in the doubled theory. This cannot be achieved by the previous
axial-vector mapping $A\leftrightarrow {\tilde A}$.
However, it is easy to see that
$$
\left(\matrix{F\cr{\tilde F}\cr}\right) \rightarrow \left(\matrix{F'\cr
{\tilde F}'\cr}\right) = \left(\matrix{0&1\cr -1 &0\cr}\right)
\left(\matrix{F\cr {\tilde F}\cr}\right)
$$
does the job. Changing variables in the lagrangian \otradoble\ to $F',
{\tilde F}'$ yields the $\mu\rightarrow -\mu$ transformation:
$$
{\cal L}\,(F, {\tilde F}, \mu) = {\cal L}\,(F', {\tilde F}', -\mu)
$$
We now perform the same gauging as before: ${\tilde F}'^{\pm}
\rightarrow {\tilde F}'^{\pm} + G^{\pm}$ and add
$$
{i\over 4\pi} \int dG* A' = {i\over 4\pi} \int (G^+ F'^+ - G^- F'^-)
$$
Finally we substitute back $F'={\tilde F}, {\tilde F}' = -F$ and find
\eqn\maaaas
{\eqalign{4\pi\, {\widetilde {\cal L}}_{\rm gauged} &= {1+\mu \over 2} (F^+
- G^+)^2 + {1+{\overline \mu}\over 2} (F^- - G^-)^2 + {1-\mu \over 2}
({\tilde F}^+)^2 + {1-{\overline\mu} \over 2} ({\tilde F}^-)^2 \cr
 &+i\mu\,(F^+ - G^+){\tilde F}^+ -i{\overline\mu}\,(F^- -G^-) {\tilde F}^- +
i(G^+ {\tilde F}^+ - G^- {\tilde F}^- )}}
And we see that we have succesfully changed the gauging from $\tilde
A$ to $A$.
Integrating out $G^{\pm}$ and gauge fixing $A= 0$ we arrive
at
\eqn\otradual
{4\pi\, {\widetilde {\cal L}} = {1-\mu \over 1+\mu} ({\tilde F}^+)^2 + {1-
{\overline\mu}\over 1+{\overline\mu}} ({\tilde F}^-)^2 }
and now the corresponding coupling ${\tilde \tau} = i {1-{\overline\mu}
 \over
1+{\overline\mu}}$ satisfies ${\tilde \tau} = -1/\tau$ as expected.

In summary, the field redefinition required in the doubled theory:
\eqn\fieldredef
{\left(\matrix{A\cr {\tilde A}\cr}\right) \rightarrow
\left(\matrix{0&1\cr (-)^p & 0\cr}\right)\left(\matrix{A\cr {\tilde
A}\cr}\right) }
is the one considered in \refs\rsenschw\ in their analysis of duality
invariant actions (note that for odd $p$ the square of this
transformation is $-1$, but this is good enough because inverting the
sign of both potentials leaves the action invariant).
However, our approach differs from that in \refs\rsenschw\ in several aspects.
Although we use an extended theory as an starting point, one of the
fields is gauged away rather than evaluated on-shell, and   we
find no conflict with Lorentz invariance. The duality transformation
is not a classical symmetry of the doubled lagrangian, yet the whole
procedure is a simple change of variables in the path integral. From
this point of view, S-duality appears as a full quantum symmetry and
perhaps one should not try to implement it classically at the
lagrangian level. Note that, in general, the effective actions are not
explicitly duality invariant because electric and magnetic variables
are mutually non-local and they never appear simultaneously in the
same low energy effective lagrangian.

\subsec{Global aspects}
 Global issues are easily dealt with in this formalism. The discrete
transformations \fieldredef\ imply that both $F$ and ${\tilde F}$ have
the same periods around homologically non-trivial $(p+1)$-manifolds.
On the other hand, the modular anomaly in arbitrary dimension requires
a more careful analysis.
On general grounds, since all path integrals are gaussian, we can
estimate the coupling constant dependence in the regularized theory as
$$
\int {DA_p \over {\rm Vol}(G_p)} e^{-S(A_p, \tau)} \sim
  ({\rm Im} \tau)^{-{1\over 2}{\rm  dim} {\cal H}'_{\rm ph}}
$$
where ${\cal H}'_{\rm ph} $ is the physical Hilbert space up to zero
modes (harmonic forms). That is, the space of $p$-forms minus the
 harmonic
ones and gauge degrees of freedom,
$$
{\rm dim} {\cal H}'_{\rm ph} = B_p - b_p -(gauge)
$$
where $b_p$ is the Betti number measuring the number of zero modes.
The pure gauge $p$-forms are $A_p = d\lambda_{p-1}$, so we can count
gauge degrees of freedom as $(p-1)$-forms, up to harmonic or exact
ones which do not really contribute to the gauge invariance of the
original theory. If we continue this nested counting until we reach
zero forms we get
$$
\eqalign{{\rm dim}{\cal H}'_{\rm ph}
 &= (B_p -b_p - (B_{p-1} - b_{p-1} -(B_{p-2}
-b_{p-2} -\,\,\, \cdots\cr
 & = (-)^p\sum_{j=0}^{p} (-)^j (B_j - b_j) \equiv {\cal
N}_p + (-)^{p+1} \sum_{j=0}^{p} (-)^j b_j \cr}
$$
 A more formal derivation of this formula, using the Fadeev-Popov
procedure, will be given in the next section. Following \refs\rwitt\ we
want to get rid of the regularization dependent factor, so that we
define the functional measure as
\eqn\witdef {
Z_p (\tau) \sim ({\rm Im}\tau)^{{1\over 2}{\cal N}_p}
\int {DA_p \over {\rm Vol}(G_p)} \,{\rm e}^{- S(A_p, \tau)} \sim ({\rm Im}
\tau) ^{{(-)^{p}\over 2}\sum_{j}^{p} (-)^j b_j}
}
 Here the measure $DA_p$ does not contain  additional powers
of ${\rm Im}\tau$,  and this prescription is  the generalization of the
sigma-model local measure of eq. \measure. From this point of view,
the modular duality anomaly is a generalization of the dilaton shift
phenomenon in two dimensional T-duality.
In obtaining \witdef\ from the coset construction, the $G^{\pm}$
integrals generate a factor
$$
(1\mp \mu)^{-{1\over 2} B_{p+1}^{+}} (1\mp {\overline \mu})^{-{1\over 2}
B_{p+1}^{-}}
$$
which must be canceled (here $B_{p+1}^{\pm}$
 are the numbers of self dual and anti-self dual
$(p+1)$-forms).
The complete coset formula for the partition function is then
$$
Z_p (\tau) = ({\rm Im}\tau)^{{1\over 2}{\cal N}_p}
\,(1-\mu)^{{1\over 2} B_{p+1}^{+}}\, (1-{\overline\mu})^{{1\over 2}
B_{p+1}^{-}}
 \,\int {DA D{\tilde A} DG \over {\rm Vol} (G_{p+1})}
\,{\rm  e}^{-
\int {\cal L}_{\rm gauged} ({\tilde F}+G, A)}
$$
and for the dual model
$$
{\widetilde Z}_p({\tilde \tau}) = ({\rm Im}{\tilde \tau})^{{1\over 2}
{\cal N}_p}
\,(1+\mu)^{{1\over 2} B_{p+1}^{+}}\, (1+{\overline\mu})^{{1\over 2}
B_{p+1}^{-}}
\,\int {DA D{\tilde A} DG \over {\rm Vol}(G_{p+1})}
\,{\rm e}^{-\int {\widetilde {\cal L}}_{\rm gauged} (F+G, {\tilde A})}
$$
where ${\tilde \tau} = -1/\tau$. Since both path integrals are
formally equal, we obtain from $\tau =i {1+{\overline \mu}
 \over 1-{\overline \mu}}$, up to
numerical constants:
$$
Z_p (\tau) = \tau^{-{1\over 2} B_{p+1}^{-}}\,\, {\overline \tau}^{-{1\over 2}
B_{p+1}^{+}}\,\, (\tau {\overline \tau})^{{1\over 2}{\cal N}_p}\,\,\,
 {\widetilde Z}_p (-1/\tau)
$$

Now, let us assume that the regularization procedure (a lattice for
example) is self dual in the sense that $B_j = B_{d-j}$ (in two
dimensions this corresponds to lattices with the same numbers of
vertices and faces). We also assume that the difference $B_{p+1}^{+} -
B_{p+1}^{-}$ is equal to $b_{p+1}^+ - b_{p+1}^- = \sigma$, the
generalized signature. Then we can write the Euler character as
$$
\chi = \sum_{j=0}^{d} (-)^j b_j =
\sum_{j=0}^{d} (-)^j B_j  = 2(-)^p {\cal N}_p + (-)^{p+1}
B_{p+1}
$$
and $B_{p+1}^{\pm} = (B_{p+1} \pm \sigma)/2$. From here we can derive
the general formula
\eqn\anomaly
{Z_p (\tau) = \tau^{-{1\over 4} \left( (-)^{p+1} \chi - \sigma
\right)}\,\,\, {\overline \tau}^{-{1\over 4}\left((-)^{p+1}\chi +\sigma\right)}
\,\,\,{\widetilde Z}_p (-1/ \tau)}
for even $p$: $(p=0, d=2)$, $(p=2, d=6)$, $(p=4, d=10)$, etc. there is
no theta term and we must take $\tau$ pure imaginary. Then the anomaly
equation reduces to
\eqn\noteta
{Z_p (g) =(\sqrt{4\pi} / g)^{\chi}\,\,\, {\widetilde Z}_p (4\pi /g)}
For odd $p$: $(p=1, d=4)$, $(p=3, d=8)$, etc. the resulting formula
looks exactly like the $d=4$ case derived in \refs\rwitt .

Regarding the possible $SL(2,{\bf Z})$ extension of the duality
symmetry for odd $p$, it depends
on the intersection matrix of $(p+1)$-forms. With our normalization of
the action, an integral shift $\tau \rightarrow \tau + n$ inserts the
term
$$
{\rm exp}\left({in\over 4\pi} \int (F_{p+1}^+)^2 - (F_{p+1}^-)^2
\right) =
{\rm exp}\left({in\over
4\pi} \int F_{p+1} \wedge F_{p+1}\right)
$$
Since $F_{p+1}$ has periods in $2\pi {\bf Z}$, the Hodge decomposition
has the form
$$
F_{p+1} = dA_p + h_{p+1} = dA_p + \sum_I 2\pi m^I \alpha_I
$$
where $m^I \in {\bf Z}$ and $\alpha_I$ are normalized harmonic forms:
$\oint_{\Sigma_I} \alpha_J = \delta^{I}_{J}$. The total shift is then
$$
{\rm exp}\left(2\pi i \, {n\over 2} m^I m^J
\int \alpha_I \wedge \alpha_J\right)
$$
Thus, as in four dimensions, we have full $SL(2,{\bf Z})$ invariance for
even intersection forms, or just $\tau \rightarrow \tau +2$ for the
general case.

\newsec{Generalized Partition functions}
In this section we discuss the general structure of partition
functions of $p$-form theories in $d$-manifolds:
\eqn\partp
{Z_p (g,\theta, J) = g^{-{\cal N}_p} \sum_{h_{p+1}}
 \int{DA_p \over {\rm Vol}(G_p)}\,\, {\rm e}^{-\int {\cal
L}_p} }
with a lagrangian
\eqn\lagp
{{\cal L}_p = {1\over g^2} F_{p+1}^2 + {i\theta \over 8\pi^2} F_{p+1}
*F_{p+1} -iJ_p^e A_p }
The theta term is only present for $d=2p+2$, $p+1={\rm even}$,
precisely in this case the dual form has rank $d-p-2=p$ and we have
the electric-magnetic case. We have also included ``electric" sources
which we assume conserved to preserve gauge invariance: $\delta J_e
=0$, where $\delta $ is the co-derivative defined as
$$
\delta = (-)^{dp +p+1} *d\,*
$$
acting on $p$ forms. The presence of a source term spoils duality, but
different choices of $J_e$ are useful to study the duality
transformations of operator insertions, like Wilson lines. Duality can
be restored by introducing monopole sources, which appear naturally in
lattice formulations as dislocations (see for example \refs\rspwil).
However, there is no elegant method to include them in continuum
treatments, due to their singular nature.

An interesting observation is that one may restore self-duality by a
suitable coupling to a smooth $(p+2)$-form. This is suggested by the
structure of the Hodge decomposition of the field strength
$$
F_{p+1} = \delta \phi_{p+2} + dA_p + h_{p+1}
$$
Normally one takes $F$ to be a closed form and drops the first term.
If we nevertheless keep it as an external source, it contributes to
$dF \neq 0$. So we can define a ``magnetic" current
\eqn\fakemag
{J_{d-p-2}^{m} = {1\over \sqrt{8\pi^2}} *d\,\delta\,\phi_{p+2} }
This is  a $p$ form precisely when a theta term is possible, but we
must stress that, since we take $\phi_{p+2}$ as a smooth form, $J_m$ is
not really a monopole current. Note however that the definition
\fakemag\ makes it automatically conserved: $\delta J^m =0$.

Since the path integral is gaussian, it has the following factorized
structure
$$
Z_p(g,\theta, J_e, J_m) = Z_{\rm source} (g,\theta, J_e, J_m)\,\,
 Z_{\rm global} (g,\theta)
\,\,Z_{\rm non-compact} (g)
$$
We now turn to a more detailed analysis or the different factors.
\subsec{Source partition function}
The source dependence is easily solved in terms of the Green function
for the corresponding laplacian, which is inverted in the space
orthogonal to the zero modes (harmonic forms).
$$
Z_{\rm source}(g,\theta, J_e, J_m) =
 {\rm exp}\left(-{g^2\over 2} J'_e {1\over
\Delta}J'_e\right)\,\, {\rm exp}\left(-{8\pi^2
 \over g^2} J_m {1\over \Delta} J_m \right)
$$
where
$$
J'_e = J_e + {\theta\over \sqrt{2\pi^2}} J_m
$$
The mixing between  electric and magnetic currents for $\theta\neq 0$
comes from a non vanishing cross term in the topological lagrangian
$F\wedge F$. This is the well known phenomenom discovered by Witten in
\refs\rwit : magnetic currents contribute to electric currents in the
presence of a theta angle.

The important point about this expression is that it is formally
self-dual under $g\rightarrow 4\pi/g$ provided we also exchange $J'_e
\leftrightarrow J_m$. However, the analogy between $J_m$ and a real
monopole current is not complete. If $J_m$ were really  monopole
currents we would have found an extra cross term coupling  $J_m$
directly to $J'_e$. This is the Aharonov-Bohm interaction resposible
for the Dirac quantization rule of electric and magnetic charges.
A more complete analysis of the physics of the source partition
function in lattice models can be found in \refs\rspwil.
\subsec{Global partition function}
The fact that the $U(1)$ group on $p$ forms is taken to be compact is
reflected (in the absence of monopoles) by the presence of quantized
fluxes around homologically non-trival $(p+1)$-submanifolds,
$\Sigma_I$. This
leads to the following generalized Theta function coming form the
classical action evaluated on harmonic forms \refs\rver
$$
Z_{\rm global}(g,\theta)=\sum_{h_{p+1}}{\rm e}^{-S_{\rm cl}} =
\sum_{m^I} {\rm e}^{-{4\pi^2\over g^2} m^I G_{IJ}m^J + {i\theta\over
2}m^I Q_{IJ}m^J}
$$
where $G_{IJ} = \int \alpha_I \wedge *\alpha_J$, $Q_{IJ} = \int
\alpha_I \wedge \alpha_J$ for normalized harmonic $(p+1)-$forms
$\oint_{\Sigma_J} \alpha_I = \delta_I^J$. The interesection matrix
$Q_{IJ}$ controls the symmetry under $\theta-$shifts. Note that
Poincare duality implies
$$
G_{IJ} = \int *\alpha_I \wedge **\alpha_J \,\,,\,\,\,\,Q_{IJ} = \int
*\alpha_I \wedge *\alpha_J
$$
So the same global partition function appears in terms of
harmonic $(d-p-1)$-forms $*\alpha_I$.

{}From the duality of the full partition function one can deduce $\it{a
\,\,posteriori}$ that the global partition function transforms under
duality as a modular form of weights $({1\over 2}
b_{p+1}^- ,{1\over 2} b_{p+1}^+)$:
$$
Z_{\rm global} (\tau) = \tau^{-{1\over 2} b_{p+1}^-} \,\,\, {\overline
\tau}^{-{1\over 2} b_{p+1}^+} \,\,\, Z_{\rm global} (-1/\tau)
$$

\subsec{Non compact partition function}
The remaining term is a standard path integral which contains the
explicit powers of the coupling $g$, and is independent of the
topological angle $\theta$. According to the heuristic arguments in
section 2 we should find
$$
Z_{\rm non-compact}(g) = g^{- {\cal N}_p} \int {DA_p
\over {\rm Vol}(G_p)} {\rm e}^{-{1\over g^2}\int (dA_p)^2} =
g^{(-)^{p+1} \sum_j^p (-)^j b_j} \times ({\rm determinants})
$$
We can derive this scaling and compute the structure of the determinant
terms by repetedly using the Fadeev-Popov trick. Let us
 introduce a gauge
fixing condition $f(A_p)$ in the first path integral
$$
\eqalign{\int{D'A_p \over {\rm Vol}(G_p)} {\rm e}^{-S(A_p)}& = \int{D'A_p \over
{\rm Vol}(G_p)}{D' \lambda_{p-1} \over {\rm Vol}(G_{p-1})}
 \Delta_{FP}
^{(p)}(A_p) f(A_p - d\lambda_{p-1}) {\rm e}^{-S(A_p)} \cr
  &= \int D'A_p \Delta_{FP}^{(p)} (A_p) f(A_p) {\rm e}^{-S(A_p)} \cr }
$$
Here $D'$ means that we do not integrate over the (finite dimensional)
space of zero modes. Also, the integral over the gauge degrees of
freedom involves as gauge ambiguity  in itself, because exact $(p-1)$
forms do not contribute to $G_p$. This is implicit in the formal
identity
$$
{\rm Vol}(G_p) = \int {D' \lambda_{p-1} \over {\rm Vol}(G_{p-1})}
$$
which we used to cancel the group volume. This means that, in
evaluating the Fadeev-Popov determinant we have to gauge-fix again:
$$
\Delta_{FP}^{(p)} (A_p)^{-1} = \int {D'\lambda_{p-1} \over {\rm
Vol}(G_{p-1})} f(A_p - d\lambda_{p-1}) = \int D' \lambda_{p-1}
\Delta_{FP}^{(p-1)} (\lambda_{p-1}) f(\lambda_{p-1}) f(A_p -
d\lambda_{p-1})
$$
where we have used
${\rm Vol}(G_{p-1}) = \int {D' \lambda_{p-2} \over {\rm
Vol}(G_{p-2})}$. The new FP functional $\Delta_{FP}^{(p-1)}
(\lambda_{p-1})$ has a similar expression as a path integral over
$\lambda_{p-2}$ forms with  $G_{p-3}$ gauge ambiguity. In this way,
the process of nested gauge fixings continues until we reach zero
forms. Using the Feynman gauge at all stages:
$$
f(\phi) = \int DC\,\, {\rm e}^{-{1\over g^2} \int C^2}\, \delta [\delta \phi
- C] = {\rm e}^{-{1\over g^2} \int (\delta \phi)^2}
$$
we find for the first determinant
 (notice that, acting on zero forms, $\delta d = \Delta_0$)
$$
\Delta_{FP}^{(1)} (\lambda_1)^{-1} = \int D' \lambda_0 \,\,{\rm
e}^{-{1\over g^2} \int (\delta \lambda_1 - \delta d \lambda_0)^2} =
g^{B_0 - b_0}\, {\rm det}'_0 (\Delta)^{-1}
$$
This is the standard result for the electromagnetic Fadeev-Popov
determinant. Now we can plug this
in the expression for the second determinant:
$$
\eqalign{\Delta_{FP}^{(2)} (\lambda_2)^{-1}& =
 \int D' \lambda_1 \,\Delta_{FP}^{(1)} (\lambda_1) \,{\rm e}^{-{1\over
g^2} \int (\delta \lambda_1)^2 + (\delta \lambda_2 - \delta d
\lambda_1)^2 } \cr
&= g^{b_0 - B_0 +B_1 -b_1}\,\, {\rm det}'_0 (\Delta)
\,\, {\rm det}'_1 (d\delta
+ (\delta d)^2)^{-1/2}\,\, {\rm exp}\left(-{1\over g^2} \int \lambda_2 K_2
\lambda_2\right) }
$$
where $K_2$ is the following operator acting on the space of two
forms:
$$
K_2 = d\delta - d\delta d\, {1\over d\delta + (\delta d)^2 }\, \delta d
\delta
$$
It is then easy to proceed and calculate all Fadeev-Popov
determinants. We find the following alternating structure:
$$
\eqalign{\Delta_{FP}^{(n)} (\lambda_n) =& g^{(b_{n-1} - B_{n-1})- (b_{n-2} -
B_{n-2}) + \cdots} \times {\rm det}'(L_{n-1})^{1/2} \,\, {\rm det}'
(L_{n-2})^{-1/2} \times \cdots \cr &\times {\rm exp}\left({1\over g^2} \int
\lambda_n K_n \lambda_n\right)}
$$
In this expression, the operators $K_n$ and
$L_n$ act on the space of $n$-forms
and are defined iteratively as
$$
\eqalign{L_n &= (\delta d)^2 + d\delta d {1\over L_{n-1}}
\delta d \delta \cr
K_n &= d\delta +(\delta d)^2 - L_n }
$$
with $L_0 = (\delta d)^2 = \Delta_0^2 $.

Finally, combining the different pieces we obtain the final result for
the partition function  in the form
\eqn\uff
{Z_{\rm non-compact} (g) = g^{(-)^{p+1} \sum_j^p (-)^j b_j}\,\,\,
{\prod_{j=0}^{p-1}
 {\rm det}'
(L_{p-j-1})^{(-)^j \over 2} \over {\rm det}' (\Delta_p - K_p)^{1\over 2} } }

\newsec{From $p$ forms to $(d-p-2)$ forms}

 The Lagrange multiplier method of ref.\refs\rwitt\ is easily
extended to the most general duality transformation between $p$-forms
and $(d-p-2)$-forms in arbitrary dimension.
We will first consider the general case without a theta term in the
lagrangian, and also drop the co-exact source term \fakemag. We
however keep the electric source to study the order-disorder mapping.

 In order to actually perform the duality transformation we need to
write all terms in first order form, that is, in terms of the field
strength. For the source term, this is easily accomplished if the
conserved current has no harmonic piece (it is purely local). Then we
can write $J_p = \delta J'_{p+1}$ and $J'$ can be chosen   to be exact
$J'_{p+1} = d J''_p $. This is only consistent if $J''$ is also
conserved $\delta J'' =0$ and has no harmonic piece (otherwise it
would lead to a vanishing $J_p$). This means that, in fact, $J_p =
\Delta_p J''_p$ and we can write
\eqn\mono {
J'_{p+1} = d\, {1\over \Delta} \, J_p
}
where, as always, the Green function of the laplacian ${1\over
\Delta}$ is defined in the space orthogonal to the harmonic forms. In
terms of $J'$ it is easy to write  the source term in first order form
as
\eqn\monon {
\int J_p A_p = \int \delta J'_{p+1} A_p = \int J'_{p+1} dA_p = \int
J'_{p+1} F_{p+1}
}
Now we are ready to introduce the fake gauge field $G$ and write the
partition function as a constrained gauge theory:
$$
\eqalign{ Z_p (g, J) &= g^{-{\cal N}_p} \sum_{h_{p+1}} \int
{DA_p \over {\rm Vol}(G_p)}\,\, {\rm e}^{-\int {\cal L} (F,J')} \cr
&= g^{-{\cal N}_p} \sum_{h_{p+1}} \int {DA_p DG \over {\rm
Vol} (G_{p+1})}\, \delta [ dG]\, {\rm e}^{-\int {\cal L}(F+G, J')}}
$$
the local delta functional includes also a periodic delta function
reducing the periods of $G$ to $2\pi \times ({\rm integer})$. In this
way, solving the constraint and gauge fixing $G$ to zero projects
$G_{p+1}$ onto  the
original gauge group $G_p$.
The constraint  is easily exponentiated in terms of a $(p+2)$-form
$$
\eqalign{\delta[dG] &= \int {D\chi_{p+2} \over {\rm Vol}(G_{p+2})}
\sum_{n^I} {\rm exp}\left( {i\over 2\pi} \int(dG\, \chi_{p+2} + 2\pi G
\,n^I \alpha_I)\right) \cr
&=\sum_{h_{p+1}} \int{D\chi_{p+2} \over {\rm Vol}(G_{p+2})}\,{\rm
exp}\left({i\over 2\pi} \int G(
\delta \chi_{p+2} + h_{p+1})\right) }
$$
and the gauge group $G_{p+2}$ appears because this representation of the
delta functional has a gauge ambiguity $\chi_{p+2} \rightarrow
\chi_{p+2} + \delta \psi_{p+3}$. Putting all the terms together and
integrating $G$ out we obtain
$$
Z_p (g,J) = g^{-{\cal N}_p} g^{B_{p+1}}\sum_{h_{p+1}}
 \int {D\chi_{p+2} \over\, {\rm
Vol}(G_{p+2})} {\rm exp}\left( -{g^2 \over 16\pi^2} \int (\delta
\chi_{p+2} + h_{p+1} + 2\pi J'_{p+1})^2 \right)
$$

We end up with a theory of $(p+2)$-forms with an ``inverted" gauge
symmetry, in terms of the co-derivative. A similar analysis to the one
performed in the previous section would reveal that the natural
definition of the gauge invariant measure in this path integral
includes a prefactor
$$
g^{(-)^p \sum_{j=p+2}^{d} (-)^j B_j}
$$
so that the modular anomaly (the net power of $g$) is given by
$g^{(-)^{p+1} \chi}$ as expected. Notice that at this point we had no
need for selfduality at the level of the regulator. However, if we
want to write the dual model as a theory of $(d-p-2)$-forms with
standard gauge invariance, then we must think of the form ${\tilde
A}_{d-p-2} = * \chi_{p+2} $ in the regulated theory
 as defined on the dual lattice.
Then, the dual partition function without sources is
\eqn\dualfinal
{\eqalign{{\widetilde Z}_{d-p-2} (4\pi/g) &=\left({ 4\pi\over g}
\right)^{-{\cal N}^*_{d-p-2}}
\sum_{{\tilde h}_{d-p-1}} \int {D{\tilde A}_{d-p-2} \over {\rm
Vol}({\tilde G}_{d-p-2})}\,\, {\rm e}^{-\int {\widetilde{\cal L}}} \cr
{\widetilde {\cal L}} &= {g^2 \over 16\pi^2} (d{\tilde A}_{d-p-2} +
{\tilde h}_{d-p-1} )^2 }}
Finally, by the definition of dual lattice, we have $B^*_j = B_{d-j}$.
In this way the powers of $g$ combine such that the modular weight
 depends only on the Euler character.
If we include the sources, we get the general duality relation
\eqn\gendul
{Z_{p} (g, J) \,=\,(\sqrt{4\pi}/ g)^{(-)^{p}\chi}\,\,
   {\widetilde Z}_{d-p-2} (4\pi
/g)_{\rm frustrated} }
where the frustrated partition function is the same as \dualfinal\
with a modified lagrangian
$$
{\widetilde {\cal L}}_{\rm frustrated} = {g^2 \over 16\pi^2} (d{\tilde
A}_{d-p-2} + {\tilde h}_{d-p-1} + 2\pi * J'_{p+1})^2
$$
These formulas are easily generalized to the case where a theta term is
present, $d-p-2=p$. The integration over the fake gauge field $G$ is
easier in terms of the selfdual and anti-selfdual projections
$G^{\pm}$ and the result is a frustrated partition function with the
modular anomaly of eq. \anomaly.

 The frustration means that we cannot absorb the $*J'$ term in a
continuous redefinition of ${\tilde A}$. This is due to the fact that,
acording to \mono, $*J'$ cannot be written as smooth exact
differential. In fact $*J'$ is co-exact and acts as a monopole
current, because it prevents the effective field strength of the dual
theory from being closed.

The most interesting sources to consider are distributions localized
on closed $p$-manifolds which lead to generalized Wilson lines. For
example, for vectors, choosing $J$ as the electromagnetic current of
first quantized particles of charges $Q_i$
$$
J = \sum_j Q_j \int_{C_j} d\tau_j {dx^{\alpha} \over d\tau_j} \,
\delta [ x-x(\tau_j)] \, dx_{\alpha}
$$
induces a term
$$
\int JA = \sum_j Q_j \oint_{C_j} A
$$
and the corresponding partition function is in fact a correlator of
Wilson lines.
$$
Z_1 (g,J) = Z_1 (g, J=0) \, \left\langle \prod_j {\rm e}^{iQ_j
\oint_{C_j} A} \right\rangle
$$

In the general case, since $\delta J=0$ is conserved and has no
harmonic piece, we have
$$
\int_{{\cal M}_d} J A = Q \,\oint_{\Sigma_p} A_p
$$
where $\Sigma_p$ is the boundary of its interior $\Sigma_p = \partial
\Sigma_p^0$. Then, by the Stokes theorem and equation \monon\
\eqn\hala
{Q \oint_{\Sigma_p} A_p = Q \int_{\Sigma_p^0} dA_p = \int_{{\cal M}_d}
J'_{p+1} dA_p }
so that $J'_{p+1}$ in this case is equal to a distribution of value
$Q$ on the $(p+1)$-dimensional ball $\Sigma_p^0$ and zero outside.
 We then see the geometrical meaning of the frustration: if we want to
write $*J'$ as an exact differential we must introduce discontinuities
across the $(p+1)$-manifold $\Sigma_p^0$. In general, such
discontinuities are better visualized in a lattice formulation, where
$p$-forms are functions over the $p$-cells of the simplicial
decomposition. Then, since ${\tilde A}_{d-p-2}$ is valuated on
$(d-p-2)$-cells of the dual lattice, the frustration amounts to a
$2\pi Q$ shift of $d{\tilde A}$   over those $(d-p-1)$-cells dual to
$\Sigma_p^0$. For $d=4$, $p=1$ this is just the 't Hooft loop
construction for compact QED.

 When $\Sigma_p$ is codimension two in
the space-time manifold, then the dual form ${\tilde A}$ is a scalar
and a simple continuum construction can be given, which generalizes
the vortex lines of two dimensional T-duality.
If $p=d-2$, then
  $\Sigma_p^0$ is a $(d-1)$-dimensional submanifold of ${\cal
M}_d$ such that $\partial ({\cal M}_d - \Sigma_p^0 ) = (\Sigma_p^0)^+
- (\Sigma_p^0)^- $, where $(\Sigma_p^0)^{\pm}$ are the ``up" and
``down'' faces as seen from ${\cal M}_d$. Returning to \hala\ we can
use the Stokes theorem on ${\cal M}_d - \Sigma_p^0$ to write
$$
\int_{\Sigma_p^0} dA_p = \int_{(\Sigma_p^0)^+} 1\cdot dA_p -
\int_{(\Sigma_p^0)^-} 0\cdot dA_p = \int_{{\cal M}_d - \Sigma_p^0}
d\alpha_{\Sigma} \wedge dA_p
$$
where $\alpha_{\Sigma} $ is a scalar function with a unit jump
$\alpha_{\Sigma^+} -\alpha_{\Sigma^-} = 1$ across $\Sigma_p^0$. Now we
can use this discontinuous function to define $*J' =
Q \, d\alpha_{\Sigma}$. This is a generalization of the two dimensional
case, where  $\Sigma_p$ is a pair of points, $\Sigma_p^0$ is the
cut joining them, and $\alpha_{\Sigma}$ is an angular variable with
respect to the cut. In this way the dual field has the boundary
conditions of a pair of oppositely charged vortices, and the partition
function is a two-point correlator of winding mode operators.

\newsec{Conclusions}

 We have presented a unified picture of the abelian S-duality of
$p$-form theories in arbitrary euclidean space-times. A simple
extension of the methods of two dimensional sigma-model duality
addresses local and global questions in the general case, including
the order-disorder mapping on gauge invariant observables, and the
modular duality anomaly, which appears as a simple generalization of
the dilaton shift in sigma model duality. Also, Ro{\v c}ek-Verlinde
coset constructions are easily generalized, leading to the same
results as the Lagrange multiplier method.

 Perhaps the most interesting open question is the existence of  a non
abelian version of the dual coset procedure.
Such a generalization would be useful in understanding the more
complicated non abelian dualities mentioned in the introduction.
It is more likely that
 a non abelian generalization of the Lagrange multiplier method
is easier to study, because the two dimensional counterpart for sigma
models is known \refs\rquevedo . Notice, however, that the auxiliary
gauge field involved in, for example, four dimensional
electric-magnetic duality, is a two-form, and it is notoriously
difficult to construct a non abelian three-form field strength with
the right properties. In general, non abelian generalizations of
higher rank gauge theories do not exist.

\newsec{Acknowledgements}
It is a pleasure to thank Y. Lozano for useful discussions.
This work was supported by NSF PHY90-21984 grant.

\listrefs
\bye